\documentclass{article}
\usepackage{amsmath}
\usepackage{amssymb}

\input{tcilatex}
\begin{document}

\begin{center}
465ClassSolvDiscrTimeDynSyst190313\bigskip

{\Huge Two Peculiar Classes of Solvable Systems Featuring 2 Dependent
Variables Evolving in Discrete-Time via 2 Nonlinearly-Coupled First-Order
Recursion Relations}

\bigskip

\textbf{Francesco Calogero}$^{a,b,1}${\LARGE \ {\large and} }\textbf{Farrin
Payandeh}$^{a,c,2}$

$^{a}$ Physics Department, University of Rome "La Sapienza", Rome, Italy

$^{b}$ INFN, Sezione di Roma 1

$^{c}$ Department of Physics, Payame Noor University (PNU), PO BOX
19395-3697 Tehran, Iran

$^{1}$ francesco.calogero@roma1.infn.it, francesco.calogero@uniroma1.it

$^{2}$ f\_payandeh@pnu.ac.ir, farrinpayandeh@yahoo.com

\bigskip

\textit{Abstract}
\end{center}

In this paper we identify certain peculiar systems of $2$ \textit{%
discrete-time} evolution equations,%
\begin{equation*}
\tilde{x}_{n}=F^{\left( n\right) }\left( x_{1},x_{2}\right) ~,~~~n=1,2~,
\end{equation*}%
which are \textit{algebraically solvable}. Here $\ell $ is the
"discrete-time" \textit{independent} variable taking \textit{integer} values
($\ell =0,1,2,...$), $x_{n}\equiv x_{n}\left( \ell \right) $ are $2$
dependent variables, and $\tilde{x}_{n}\equiv x_{n}\left( \ell +1\right) $
are the corresponding $2$ \textit{updated} variables. In a previous paper
the $2$ functions $F^{\left( n\right) }\left( x_{1},x_{2}\right) ,$ $n=1,2,$
were defined as follows: $F^{\left( n\right) }\left( x_{1},x_{2}\right)
=P_{2}\left( x_{n},x_{n+1}\right) ,$ $n=1,2~\func{mod}[2],$ with $%
P_{2}\left( x_{1},x_{2}\right) $ a specific \textit{second-degree
homogeneous polynomials} in the $2$ (\textit{indistinguishable}!) dependent
variables $x_{1}\left( \ell \right) $ and $x_{2}\left( \ell \right) $. In
the present paper we further clarify some aspects of that model and we
present its extension to the case when $F^{\left( n\right) }\left(
x_{1},x_{2}\right) =Q_{k}^{\left( n\right) }\left( x_{1},x_{2}\right)
,~n=1,2,$ with $Q_{k}^{\left( n\right) }\left( x_{1},x_{2}\right) $ a
specific \textit{homogeneous function of arbitrary} (\textit{integer})
\textit{degree} $k$ (hence a \textit{polynomial of degree }$k$ when $k>0$)
in the $2$ dependent variables $x_{1}\left( \ell \right) $ and $x_{2}\left(
\ell \right) $.

\bigskip

\section{Introduction and main results}

The results reported in this paper are a nontrivial extension of those
reported in \cite{CP2019a}, to which the interested reader is referred: (i)
for a terse overview of an old (see \cite{C1978} \cite{C2001})---and
recently substantially improved (see \cite{C2016} \cite{C2018} \cite{BC2018}
\cite{B2018} \cite{CP2019} \cite{CP2018} \cite{CP2019b})---technique to
identify \textit{solvable} dynamical systems in \textit{continuous-time} $t$
; (ii) for an introduction to the extension of that approach to the case of
\textit{discrete-time} $\ell $ (see \cite{BCL2014} \cite{BC2016} \cite%
{BC2017}); (iii) for a very terse review of previous results on \textit{%
analogous solvable discrete-time} models (see \cite{C2018}). To make the
relevance of the present paper immediately clear we report already in this
introductory section what we consider its \textit{main} findings.

\textbf{Notation 1-1}. Hereafter $\ell =0,1,2,$... denotes the \textit{%
discrete-time} \textit{independent} variable; the \textit{dependent}
variables are $x_{n}\equiv x_{n}\left( \ell \right) $ (with $n=1,2$), and
the notation $\tilde{x}_{n}\equiv x_{n}\left( \ell +1\right) $ indicates the
once-updated values of these variables. We shall also use other dependent
variables, for instance $y_{m}\equiv y_{m}\left( \ell \right) $ (with $m=1,2$%
) and then of course likewise $\tilde{y}_{m}\equiv y_{m}\left( \ell
+1\right) $. Variables such as $x_{n}$ and $y_{m}$ are generally assumed to
be \textit{complex} numbers (this does not exclude that they might in some
cases take only \textit{real }values); note that while these quantities
generally depend on the \textit{discrete-time} variable $\ell $, only
occasionally this is \textit{explicitly} indicated. Parameters such as $a$, $%
b$, $\alpha ,$ $\beta ,$ $\gamma ,$ $B_{n}$, $C_{j}$ (with $n=1,2$; $j=1,2,3$%
) are generally time-independent \textit{complex} numbers, while $k,$ $q,$ $r
$ are \textit{real integers}; and \textit{indices} such as $n$, $m,$ $j$ are
of course \textit{positive integers} (the values they may take shall be
explicitly indicated or be quite clear from the context). The quantity $S$
denotes an \textit{arbitrarily} \textit{assigned} sign, $S=\pm $: note that
generally the assignment of the sign $S$ may depend on the \textit{%
discrete-time} $\ell $, $S\equiv S\left( \ell \right) $ (but of course it
has the same assigned value $S\left( \ell \right) $ for each value of $\ell $%
). Finally: the convention is hereafter adopted according to which $%
\tsum\limits_{s=s_{-}}^{s_{+}}f\left( s\right) =0$ and $\tprod%
\limits_{s=s_{-}}^{s_{+}}f\left( s\right) =1$ whenever $s_{+}<s_{-}$.
\textbf{$\blacksquare $}

\textbf{Remark 1-1}. In this paper the term \textit{solvable} generally
characterizes systems of \textit{discrete-time} evolution equations the
initial-values problem of which is \textit{explicitly solvable by algebraic
operations}. \textbf{$\blacksquare $}

The\textit{\ }main result of \cite{CP2019a} is to provide the \textit{%
explicit} solution of the initial-values problem for the system of $2$
nonlinearly-coupled \textit{discrete-time} evolution equations
\begin{equation}
\tilde{x}_{n}=-\left( x_{1}+x_{2}\right) \left[ a\left( x_{1}+x_{2}\right)
+Sb\left( x_{n}-x_{n+1}\right) \right] ,~~~n=1,2~\func{mod}[2]~.
\label{1xndotuEq1}
\end{equation}%
(Note here the notational changes with respect to \cite{CP2019a}: implying $%
\alpha =2a,$ $\beta =2b$ and the \textit{explicit} introduction of the
\textit{arbitrary} $\ell $-dependent sign $S\equiv S\left( \ell \right) ,$
which is indeed implicit in the formulas (2.11b) and (1.2a) of \cite{CP2019a}%
).

\textbf{Remark 1.2}. The characteristic of the \textit{discrete-time}
evolution of this model is that, if the sign $S\left( \ell \right) $ is
\textit{positive, }$S\left( \ell \right) =+,$ then
\begin{subequations}
\label{xntildeOld}
\begin{eqnarray}
&&\tilde{x}_{n}\equiv x_{n}\left( \ell +1\right)  \notag \\
&=&-\left[ x_{1}\left( \ell \right) +x_{2}\left( \ell \right) \right]
\left\{ a\left[ x_{1}\left( \ell \right) +x_{2}\left( \ell \right) \right] +b%
\left[ x_{n}\left( \ell \right) -x_{n+1}\left( \ell \right) \right] \right\}
,  \notag \\
&&n=1,2~\func{mod}[2]~;
\end{eqnarray}%
while if $S\left( \ell \right) =-,$ then%
\begin{eqnarray}
&&\tilde{x}_{n}\equiv x_{n}\left( \ell +1\right)  \notag \\
&=&-\left[ x_{1}\left( \ell \right) +x_{2}\left( \ell \right) \right]
\left\{ a\left[ x_{1}\left( \ell \right) +x_{2}\left( \ell \right) \right] -b%
\left[ x_{n}\left( \ell \right) -x_{n+1}\left( \ell \right) \right] \right\}
,  \notag \\
&&n=1,2~\func{mod}[2]~.
\end{eqnarray}%
So it might appear that we are dealing here with a large plurality of
\textit{distinct} dynamical systems, as yielded by all the possible ($\ell $%
-dependent!) assignments of the values---positive or negative sign---of $%
S\left( \ell \right) $. But this is not really the case, because it is
evident that the different outcomes of these two systems (\ref{xntildeOld})
is merely to exchange the roles of the variables $x_{1}\left( \ell \right) $
and $x_{2}\left( \ell \right) $; hence the system (\ref{1xndotuEq1}) yields
a well-defined, \textit{unique} evolution if we consider the two variables $%
x_{1}\left( \ell \right) $ and $x_{2}\left( \ell \right) $ to identify $2$
\textit{indistinguishable} entities, such as the $2$ different \textit{zeros}
of a generic second-degree polynomial. And it was indeed shown in \cite%
{CP2019a} that the solution of the initial-values problem for the \textit{%
discrete-time} evolution (\ref{1xndotuEq1})---with the sign $S\left( \ell
\right) $ being \textit{arbitrarily} assigned for every value of $\ell $%
---is provided by the $2$ zeros $x_{1}\left( \ell \right) $ and $x_{2}\left(
\ell \right) $ of a specific second-degree polynomial,
\end{subequations}
\begin{equation}
p_{2}\left( z;\ell \right) =z^{2}+y_{1}\left( \ell \right) z+y_{2}\left(
\ell \right) =\left[ z-x_{1}\left( \ell \right) \right] \left[ z-x_{2}\left(
\ell \right) \right] ~,
\end{equation}%
the $2$ coefficients of which, $y_{1}\left( \ell \right) $ and $y_{2}\left(
\ell \right) ,$ are \textit{unambiguously} determined and indeed \textit{%
explicitly} known in terms of the initial values $x_{1}\left( 0\right) $ and
$x_{2}\left( 0\right) $.

This remark is introduced here as an introduction to the somewhat more
peculiar phenomenon associated with the more general models considered in
the present paper, see below. \textbf{$\blacksquare $}

The \textit{first main} result of the present paper is to provide (in the
following \textbf{Section 2}) the \textit{explicit} solution of the
initial-values problem for the following, more general, system of $2$
nonlinearly-coupled \textit{discrete-time} evolution equations
\begin{subequations}
\label{1x12tilde}
\begin{equation}
\tilde{x}_{n}=\left( B_{1}x_{1}+B_{2}x_{2}\right) ^{k}\left(
E_{n1}x_{1}+E_{n2}x_{2}\right) ~,~~~n=1,2~,  \label{1x1tilde}
\end{equation}%
\begin{equation}
E_{11}=d\left( \alpha g_{1}B_{1}+S\beta B_{2}g_{2}\right) ~,~\
~E_{12}=dB_{2}\left( \alpha g_{1}+S\beta g_{3}\right) ~,
\end{equation}%
\begin{equation}
E_{21}=\alpha \left( 1-B_{1}dg_{1}\right) \left( B_{1}/B_{2}\right)
-B_{1}dS\beta g_{2}~,
\end{equation}%
\begin{equation}
E_{22}=\alpha \left( 1-B_{1}dg_{1}\right) -B_{1}dS\beta g_{3}~.
\end{equation}%
\begin{equation}
d=\left( 1/2\right) \left[ \left( B_{1}\right) ^{2}C_{2}+\left( B_{2}\right)
^{2}C_{1}-B_{1}B_{2}C_{3}\right] ^{-1}~,  \label{1d}
\end{equation}%
\begin{equation}
g_{1}=2B_{1}C_{2}-B_{2}C_{3}~,~~~g_{2}=2B_{2}C_{1}-B_{1}C_{3}~,~~~g_{3}=B_{2}C_{3}-2B_{1}C_{2}~,
\label{1g123}
\end{equation}%
with $k$ an \textit{arbitrary }(\textit{time-independent}, \textit{integer})
parameter (of course $k$ must be chosen to be a \textit{positive integer} if
one prefers that the right-hand side of these equations be homogeneous
\textit{polynomials} of degree $k+1$). The $7$ parameters $\alpha ,$ $\beta $
and $B_{n}$, $C_{j}$ (with $n=1,2$; $j=1,2,3$) are as well \textit{arbitrary}
(see \textbf{Notation 1.1}).

\textbf{Remark 1.3}. The restriction to \textit{integer} values of the
parameter $k$---and of the analogous exponents $q$ and $r,$ see below and
\textbf{Notation 1.1}---is to make sure that the right-hand sides of the
\textit{main discrete-time} evolution equations we introduce and discuss in
this paper are \textit{analytic}, hence \textit{unambiguously} defined,
functions. \textbf{$\blacksquare $}

For $k=1$ the system (\ref{1x12tilde}) can of course be re-written as
follows:
\end{subequations}
\begin{subequations}
\begin{equation}
\tilde{x}_{n}=\left( B_{1}x_{1}+B_{2}x_{2}\right) \left(
E_{n1}x_{1}+E_{n2}x_{2}\right) ~,~~~n=1,2~;  \label{1xntilde}
\end{equation}%
\begin{equation}
E_{11}=d\left( \alpha g_{1}B_{1}+S\beta B_{2}g_{2}\right) ~,~\
~E_{12}=d\left( \alpha g_{1}B_{2}+S\beta B_{2}g_{3}\right) ~,
\end{equation}%
\begin{equation}
E_{21}=\alpha \left( 1-B_{1}dg_{1}\right) \left( B_{1}/B_{2}\right)
-B_{1}dS\beta g_{2}~,
\end{equation}%
\begin{equation}
E_{22}=\alpha \left( 1-B_{1}dg_{1}\right) -B_{1}dS\beta g_{3}~.
\end{equation}

For $k=-1$ the system (\ref{1x12tilde}) can of course be re-written as
follows:
\end{subequations}
\begin{subequations}
\begin{equation}
\tilde{x}_{n}=\frac{D_{n1}x_{1}+D_{n2}x_{2}}{B_{1}x_{1}+B_{2}x_{2}}%
~,~~~n=1,2~;
\end{equation}%
\begin{equation}
D_{11}=d\left( \alpha g_{1}B_{1}+S\beta B_{2}g_{2}\right)
~,~~~D_{12}=d\left( \alpha g_{1}B_{2}+S\beta B_{2}g_{3}\right) ~,
\end{equation}%
\begin{equation}
D_{21}=\alpha \left( 1-B_{1}dg_{1}\right) \left( B_{1}/B_{2}\right)
-B_{1}dS\beta g_{2}~,
\end{equation}%
\begin{equation}
D_{22}=\alpha \left( 1-B_{1}dg_{1}\right) -B_{1}dS\beta g_{3}~.
\end{equation}

\textbf{Remark 1.4}. The case with $k=1$ is sufficiently interesting to
deserve this additional remark. Its equations of motion (\ref{1xntilde}) can
of course be reformulated as follows:
\end{subequations}
\begin{equation}
\tilde{x}_{n}=a_{n1}\left( x_{1}\right) ^{2}+a_{n2}\left( x_{2}\right)
^{2}+a_{n3}x_{1}x_{2}~,~~~n=1,2~,
\end{equation}%
with easily obtainable expressions of the $6$ parameters $a_{nj}$ ($n=1,2;$ $%
j=1,2,3$) in terms of the $7$ arbitrary parameters $\alpha ,$ $\beta $ and $%
B_{n}$, $C_{j}$ (with $n=1,2$; $j=1,2,3$; see (\ref{1x12tilde})). But this
does \textit{not} imply that these $6$ parameters $a_{nj}$ can be \textit{%
arbitrarily} assigned: the fact that the right-hand sides of the $2$
recursions (\ref{1x1tilde}) feature a \textit{common zero}---they both
vanish when $B_{1}x_{1}+B_{2}x_{2}$ vanishes---is easily seen to imply that
these $6$ parameters are constrained to satisfy (at least!) the following
nonlinear relationship:%
\begin{equation}
\left( a_{11}a_{22}-a_{21}a_{12}\right) ^{2}+\left(
a_{13}a_{21}-a_{11}a_{23}\right) \left( a_{13}a_{22}-a_{12}a_{23}\right) =0
\label{Conda}
\end{equation}%
(see, if need be, \textbf{Remark 5.3} of Ref. \cite{CP2019}). \textbf{$%
\blacksquare $}

The \textit{second main} result of the present paper is to provide the
\textit{explicit} solution of the initial-values problem for the system of $%
2 $ nonlinearly-coupled \textit{discrete-time} evolution equations
\begin{equation}
\tilde{x}_{n}=\left( 2x_{1}+x_{2}\right) ^{k}\left[ \left( -1\right)
^{k}a\left( 2x_{1}+x_{2}\right) +S\left( -1\right) ^{n}nb\left(
x_{1}-x_{2}\right) \right] ~,~~~n=1,2~.  \label{1xndotk}
\end{equation}%
Note that in this case---differently from those reported above---the \textit{%
discrete-time} evolutions of the $2$ variables $x_{1}\left( \ell \right) $
and $x_{2}\left( \ell \right) $ are \textit{different}; hence these
dependent variables are \textit{no more} related to each other by just an
exchange of their identities. Yet these evolution equations, (\ref{1xndotk}%
), still have a somewhat analogous property to those discussed above: for
any given pair of initial data, $x_{1}\left( 0\right) $ and $x_{2}\left(
0\right) $, only $2$ different solutions, say the two pairs $x_{1}^{\left(
+\right) }\left( \ell \right) $, $x_{2}^{(+)}\left( \ell \right) $ and $%
x_{1}^{\left( -\right) }\left( \ell \right) $, $x_{2}^{(-)}\left( \ell
\right) ,$ emerge, not $2^{\ell }$ as it might instead be inferred due to
the indeterminacy of the signs $S\left( \ell \right) $ appearing in these
evolution equations (\ref{1xndotk}) at every step of the \textit{%
discrete-time} evolution they yield. This is demonstrated by the explicit
solution of the initial-values problem for this model, as reported below;
but the interested reader may readily understand the origin of this
remarkable phenomenon by noting that a simple iteration of (\ref{1xndotk})
entails the (double-step) formula
\begin{eqnarray}
&&x_{n}\left( \ell +2\right) =\left( -1\right) ^{k}3^{k+1}a^{k}\left[
2x_{1}\left( \ell \right) +x_{2}\left( \ell \right) \right] ^{k\left(
k+2\right) }\cdot  \notag \\
&&\cdot \left\{ a^{2}\left[ 2x_{1}\left( \ell \right) +x_{2}\left( \ell
\right) \right] -S\left( \ell \right) S\left( \ell +1\right) \left(
-1\right) ^{n}nb^{2}\left[ x_{1}\left( \ell \right) -x_{2}\left( \ell
\right) \right] \right\} ~,  \notag \\
&&n=1,2~;
\end{eqnarray}%
indeed the quantity $S\left( \ell \right) S\left( \ell +1\right) $ is again
just a sign, i. e. it can only take the $2$ values $+1$ or $-1$, as implied
by the very definition of $S\left( \ell \right) ,$ see \textbf{Notation 1.1}%
. Hence this formula clearly shows that---starting from the initial values $%
x_{n}\left( 0\right) $---also at the $\ell =2$ level (as at the $\ell =1$
level)---this evolution yields \textit{only two} (\textit{not four}!)
alternative values for the pair $x_{1}\left( 2\right) ,$ $x_{2}\left(
2\right) $, say $x_{1}^{\left( +\right) }\left( 2\right) ,$ $x_{2}^{\left(
+\right) }\left( 2\right) $ (corresponding to $S\left( 0\right) S\left(
1\right) =+$) respectively $x_{1}^{\left( -\right) }\left( 2\right) ,$ $%
x_{2}^{\left( -\right) }\left( 2\right) $ (corresponding to $S\left(
0\right) S\left( 1\right) =-$); and this phenomenology of course
prevails---see below---at every subsequent level $\ell >2$ of the \textit{%
discrete-time} evolution.

Additional results and proofs---including the \textit{explicit} solutions of
the initial-values problems for the $2$ systems of $2$ \textit{discrete-time}
evolution equations (\ref{1x12tilde}) respectively (\ref{1xndotk})---are
provided in \textbf{Section 2.} A concluding \textbf{Section 3} outlines
additional developments.

\bigskip

\section{Additional results and proofs}

The starting point of our treatment in this \textbf{Section 2} is the
following \textit{algebraically solvable} system of $2$ \textit{discrete-time%
} evolution equations in the $2$ dependent variables $y_{1}\equiv
y_{1}\left( \ell \right) $ and $y_{2}\equiv y_{2}\left( \ell \right) $:%
\begin{equation}
\tilde{y}_{1}=\alpha \left( y_{1}\right) ^{1+k}~,~~~\tilde{y}_{2}=\beta
^{2}y_{2}\left( y_{1}\right) ^{q}+\gamma \left( y_{1}\right) ^{r}~,
\label{2y12tilde}
\end{equation}%
where the $6$ parameters $\alpha ,\beta ,$ $\gamma ,$ $k,$ $q$, $r$ can be%
\textit{\ a priori arbitrarily assigned} (see \textbf{Notation 1.1}; but see
also below for eventual restrictions on these parameters).

The fact that the initial-values problem for this system of $2$ \textit{%
discrete-time} evolution equations is \textit{solvable} is demonstrated by
exhibiting its solution:
\begin{subequations}
\label{2y12}
\begin{equation}
y_{1}\left( \ell \right) =\alpha ^{\left[ \left( 1+k\right) ^{\ell }-1\right]
/k}\left[ y_{1}\left( 0\right) \right] ^{\left( 1+k\right) ^{\ell }}~,
\label{2y1}
\end{equation}%
\begin{equation}
y_{2}\left( \ell \right) =\beta ^{2\ell }\alpha ^{\left( q/k^{2}\right)
\left[ \left( 1+k\right) ^{\ell }-k\ell -1\right] }\left[ y_{1}\left(
0\right) \right] ^{\left( q/k\right) \left[ \left( 1+k\right) ^{\ell }-1%
\right] }Y_{2}\left( \ell \right) ~,  \label{2y1b}
\end{equation}%
\begin{eqnarray}
&&Y_{2}\left( \ell \right) =y_{2}\left( 0\right)  \notag \\
&&+\gamma \sum_{s=0}^{\ell -1}\left\{ \beta ^{-2\left( s+1\right) }\alpha
^{u \left[ \left( 1+k\right) ^{s}-1\right] /k^{2}+\left( q/k\right) s}\left[
y_{1}\left( 0\right) \right] ^{\left[ u\left( 1+k\right) ^{s}+q\right]
/k}\right\} ~,  \label{2y1c}
\end{eqnarray}%
\begin{equation}
u\equiv kr-\left( 1+k\right) q~.  \label{2y1d}
\end{equation}%
The interested reader will verify that this solution is consistent with the
initial data $y_{1}\left( 0\right) ,$ $y_{2}\left( 0\right) $ and that it
does satisfy the system of evolution equations (\ref{2y12tilde}). Note that
the assumption that the 3 parameters $k,$ $q,$ $r$ be \textit{integers} (see
\textbf{Notation 1.1}) is \textit{necessary and sufficient} to guarantee
that all exponents in these formulas are \textit{integers}, thereby
excluding any nonanalyticity/indeterminacy in the equations of motion (\ref%
{2y12tilde}) and in their solutions (\ref{2y12}).

For reasons that shall be clear in the following, we are also interested in
the solution of the system (\ref{2y12tilde}) in the particular case when the
$2$ parameters $q$ and $r$ are expressed in terms of $k$ as follows:
\end{subequations}
\begin{equation}
q=2k~,~~~r=2\left( 1+k\right) ~.  \label{2qrk}
\end{equation}%
Note that this assignment implies $u=0$ (see (\ref{2y1d})). In this case the
sum in the right-hand side of (\ref{2y1c}) becomes a \textit{geometric} sum,
hence it can be performed \textit{explicitly}; therefore in this special
case the formulas (\ref{2y12}) are replaced by the following, \textit{more
explicit}, versions:
\begin{subequations}
\label{2y12Spec}
\begin{equation}
y_{1}\left( \ell \right) =\alpha ^{\left[ \left( 1+k\right) ^{\ell }-1\right]
/k}\left[ y_{1}\left( 0\right) \right] ^{\left( 1+k\right) ^{\ell }}~,
\label{2y1Spec}
\end{equation}%
\begin{eqnarray}
y_{2}\left( \ell \right) &=&\beta ^{2\ell }\alpha ^{2\left[ \left(
1+k\right) ^{\ell }-k\ell -1\right] /k}\left[ y_{1}\left( 0\right) \right]
^{2\left[ \left( 1+k\right) ^{\ell }-1\right] }Y_{2}\left( \ell \right) ~,
\label{2y2Specb} \\
Y_{2}\left( \ell \right) &=&y_{2}\left( 0\right) +\gamma \beta ^{-2}\left[
y_{1}\left( 0\right) \right] ^{2}\left[ \frac{\left( \alpha /\beta \right)
^{2\ell }-1}{\left( \alpha /\beta \right) ^{2}-1}\right] ~.  \label{2y2Specc}
\end{eqnarray}

Our next task is to identify various systems---satisfied by $2$ new
dependent variables $x_{1}\equiv x_{1}\left( \ell \right) $ and $x_{2}\equiv
x_{2}\left( \ell \right) $---the solution of which can be identified via the
solution of the system (\ref{2y12tilde}). To this end we set, to begin with,
\end{subequations}
\begin{subequations}
\begin{equation}
y_{1}=-\left( x_{1}+x_{2}\right) ~,~~~y_{2}=x_{1}x_{2}~,  \label{y12x12}
\end{equation}%
which clearly implies that $x_{1}$ and $x_{2}$ are the $2$ zeros of the
following second-degree monic polynomial:%
\begin{equation}
z^{2}+y_{1}z+y_{2}=\left( z-x_{1}\right) \left( z-x_{2}\right) ~,
\label{Poly1y2x1x2}
\end{equation}%
implying%
\begin{equation}
\left( x_{n}\right) ^{2}+y_{1}x_{n}+y_{2}=0~,~\hspace{0in}~n=1,2~,
\end{equation}%
\begin{equation}
x_{n}\left( \ell \right) =\left( 1/2\right) \left\{ -y_{1}\left( \ell
\right) +\left( -1\right) ^{n}\left\{ \left[ y_{1}\left( \ell \right) \right]
^{2}-4y_{2}\left( \ell \right) \right\} ^{1/2}\right\} ~,~\hspace{0in}%
~n=1,2~.  \label{1xnel}
\end{equation}%
Likewise (replacing $\ell $ with $\ell +1$)
\end{subequations}
\begin{subequations}
\begin{equation}
\left( \tilde{x}_{n}\right) ^{2}+\tilde{y}_{1}\tilde{x}_{n}+\tilde{y}%
_{2}=0~,~~~n=1,2~,
\end{equation}%
\begin{equation}
\tilde{x}_{n}=(1/2\left[ -\tilde{y}_{1}+\left( -1\right) ^{n}\tilde{\Delta}%
_{1}\right] ~,~~~\left( \tilde{\Delta}_{1}\right) ^{2}=\left( \tilde{y}%
_{1}\right) ^{2}-4\tilde{y}_{2}~.  \label{xntilde}
\end{equation}%
We then use the evolution equations (\ref{2y12tilde}) to express, in the
right-hand sides of (\ref{xntilde}), $\tilde{y}_{1}$ and $\tilde{y}_{2}$ in
terms of $y_{1}$ and $y_{2}$, and then the relations (\ref{y12x12}) to
express $y_{1}$ and $y_{2}$ in terms of $x_{1}$ and $x_{2}$; thereby getting
the following system of \textit{discrete-time} evolution equations for the $%
2 $ dependent variables $x_{1}\left( \ell \right) $ and $x_{2}\left( \ell
\right) $:
\end{subequations}
\begin{subequations}
\label{2EvEq1}
\begin{equation}
\tilde{x}_{n}=\left( 1/2\right) \left\{ -\alpha \left[ -\left(
x_{1}+x_{2}\right) \right] ^{k+1}+\left( -1\right) ^{n}\Delta _{1}\right\}
~,~~~n=1,2~,  \label{2xntilde}
\end{equation}%
\begin{equation}
\Delta _{1}=S\left\{ \alpha ^{2}\left[ -\left( x_{1}+x_{2}\right) \right]
^{2\left( k+1\right) }-4\beta ^{2}x_{1}x_{2}\left[ -\left(
x_{1}+x_{2}\right) \right] ^{q}-4\gamma \left[ -\left( x_{1}+x_{2}\right) %
\right] ^{r}\right\} ^{1/2}~.  \label{2delta1}
\end{equation}

\textbf{Remark 2.1}. The $\pm $ sign $S\equiv S\left( \ell \right) $ in this
definition (\ref{2delta1}) of $\Delta _{1}$ might be considered pleonastic
in view of the sign indeterminacy of the square-root in the right-hand side
of this formula (\ref{2delta1}). We did put it there as a reminder of the
fact that, for every value of the discrete time $\ell $, the assignment of
the labels $1$ or $2$ to the \textit{solutions} of the second-degree
evolution equation (\ref{2EvEq1}) is \textit{optional}. Indeed the two
variables $x_{1}\left( \ell \right) $ and $x_{2}\left( \ell \right) $---the
\textit{discrete-time} evolution of which is identified with the evolution
of the $2$ \textit{zeros} of the second-degree $\ell $-dependent (monic)
polynomial (\ref{Poly1y2x1x2}) the $2$ \textit{coefficients} $y_{1}\left(
\ell \right) $ and $y_{2}\left( \ell \right) $ of which evolve according to
the \textit{solvable} system (\ref{2y12tilde})---should be considered
\textit{indistinguishable}. Note that this implies that this evolution
equation is actually \textit{not quite deterministic}; it is only
deterministic for the couple of \textit{indistinguishable }dependent
variables $x_{1}\equiv x_{1}\left( \ell \right) $, $x_{2}\equiv x_{2}\left(
\ell \right) $: a well-known phenomenon for this kind of evolution
equations, as discussed above and in the past---see for instance \cite%
{BC2017} and \textbf{Chapter 7} ("Discrete Time") of the book \cite{C2018}
(in particular \textbf{Remark 7.1.2} there). \textbf{$\blacksquare $}

\textbf{Remark 2.2}. Of course when $s$ is an \textit{integer} the power $%
\left( -z\right) ^{s}$ can be replaced by $z$ respectively $-z$ for $s$
\textit{even} respectively \textit{odd}. \textbf{$\blacksquare $}

The results obtained so far allow to formulate the following

\textbf{Proposition 2.1}. The solution of the initial-values problem for the
system of \textit{discrete-time} evolution equations (\ref{2EvEq1}) is
provided---up to the limitations implied by \textbf{Remark 2.1}---by the $2$
\textit{zeros} $x_{1}\equiv x_{1}\left( \ell \right) $ and $x_{2}\equiv
x_{2}\left( \ell \right) $ of the polynomial (\ref{Poly1y2x1x2}) (see (\ref%
{1xnel})), with its \textit{coefficients} $y_{1}\equiv y_{1}\left( \ell
\right) $ and $y_{2}\equiv y_{2}\left( \ell \right) $ given by the formulas (%
\ref{2y12}) where of course (see (\ref{y12x12})) $y_{1}\left( 0\right)
=-x_{1}\left( 0\right) -x_{2}\left( 0\right) $ and $y_{2}\left( 0\right)
=x_{1}\left( 0\right) x_{2}\left( 0\right) $. \textbf{$\blacksquare $}

We believe that the interest---both theoretical and applicative---of the
system (\ref{2EvEq1}) is modest, due to the appearance of a square root in
the right-hand side of its equations of motion. Hence our next step is to
restrict attention to the values identified by eq. (\ref{2qrk}). Indeed
these assignments---beside allowing the more explicit solution of the system
of recursions (\ref{2y12tilde}) characterizing the \textit{discrete-time}
evolution of the $2$ dependent variables $y_{1}\left( \ell \right) $ and $%
y_{2}\left( \ell \right) ,$ see (\ref{2y12Spec})---also allow (remarkably!)
to get rid of the square-root in the right-hand side of (\ref{2delta1}),
provided we moreover make the assignments
\end{subequations}
\begin{equation}
\gamma =a^{2}-b^{2}~,~~~\alpha =2a~,~~~\beta =2b~;  \label{gammaCase1}
\end{equation}%
obtaining thereby just the system of evolution equations (\ref{1x12tilde}).
This allows us to prove our \textit{first main} result, in the guise of the
following

\textbf{Proposition 2.2}. The solution of the initial-values problem for the
system of \textit{discrete-time} evolution equations (\ref{1x12tilde}) is
provided by the $2$ \textit{zeros} $x_{1}\equiv x_{1}\left( \ell \right) $
and $x_{2}\equiv x_{2}\left( \ell \right) $ of the polynomial (\ref%
{Poly1y2x1x2}) (see (\ref{1xnel})), with its \textit{coefficients} $%
y_{1}\equiv y_{1}\left( \ell \right) $ and $y_{2}\equiv y_{2}\left( \ell
\right) $ given by the formulas (\ref{2y12Spec}) (with the assignments (\ref%
{gammaCase1})), where of course (see (\ref{y12x12})) $y_{1}\left( 0\right)
=-x_{1}\left( 0\right) -x_{2}\left( 0\right) $ and $y_{2}\left( 0\right)
=x_{1}\left( 0\right) x_{2}\left( 0\right) $.

Let us again emphasize that, for each value of $\ell $, the assignment of
the labels $1$ or $2$ to the two zeros of the polynomial (\ref{Poly1y2x1x2})
is \textit{optional}. \textbf{$\blacksquare $\ }

This \textbf{Proposition 2.2 }corresponds to the \textit{first main} result
reported in \textbf{Section 1}.

Our next step is to modify the relationship among the variables $x_{n}\left(
\ell \right) $ and $y_{n}\left( \ell \right) $ by replacing the \textit{%
second-degree} (monic) polynomial (\ref{Poly1y2x1x2}) with the following
(monic) \textit{third-degree} polynomial:
\begin{subequations}
\label{2Pol3}
\begin{equation}
z^{3}+y_{1}z^{2}+y_{2}z+y_{3}=\left( z-x_{1}\right) ^{2}\left(
z-x_{2}\right) ~,  \label{Pol3}
\end{equation}%
where of course now%
\begin{equation}
y_{1}=-\left( 2x_{1}+x_{2}\right) ~,~~~y_{2}=x_{1}\left( x_{1}+2x_{2}\right)
~,~~~y_{3}=-\left( x_{1}\right) ^{2}x_{2}~.  \label{y123x12}
\end{equation}%
Note that---following \cite{BC2018} and \cite{CP2019}---we are now focussing
on a \textit{special} polynomial of degree $3$ which features only $2$
\textit{zeros}, the \textit{zero} $x_{1}$ with multiplicity $2$ and the
\textit{zero} $x_{2}$ with multiplicity $1$: this of course implies that
these $2$ \textit{zeros} are now quite \textit{distinct}, and moreover that
now only $2$ of the $3$ \textit{coefficients} $y_{1},$ $y_{2},$ $y_{3}$ can
be arbitrarily assigned, the unassigned one being determined in terms of the
other two by the requirement that the $3$ equations (\ref{y123x12}) be
simultaneously satisfied.

\textbf{Remark 2.3}. It is for instance easy to check that the \textit{%
coefficient} $y_{3}$ of the polynomial (\ref{Pol3}) is given by the
following formula in terms of the other $2$ \textit{coefficients} $y_{1}$
and $y_{2}$:
\end{subequations}
\begin{subequations}
\begin{equation}
y_{3}=\left\{ -2\left( y_{1}\right) ^{3}+9y_{1}y_{2}+2S\left[ \left(
y_{1}\right) ^{2}-3y_{2}\right] ^{3/2}\right\} /27~.
\end{equation}%
Likewise, from the first $2$ of the $3$ formulas (\ref{y123x12}) one easily
obtains the following expressions of the two zeros $x_{1}$ and $x_{2}$ in
terms of the $2$ coefficients $y_{1}$ and $y_{2}$:
\begin{equation}
x_{n}=\left( 1/2\right) \left\{ -y_{1}+S\left( -1\right) ^{n}n\left[ \left(
y_{1}\right) ^{2}-3y_{2}\right] ^{1/2}\right\} ~,~~~n=1,2~.
\label{2Solx12y12}
\end{equation}%
Note the ambiguity in these formulas implied by the indeterminacy of the
sign $S$, see \textbf{Notation 1.1}. \textbf{$\blacksquare $}

It is now convenient to write again these expressions of the $2$ \textit{%
zeros} $x_{n}\equiv x_{n}\left( \ell \right) ,$ but with $\ell $ replaced by
$\ell +1$:
\end{subequations}
\begin{equation}
\tilde{x}_{n}=(1/2)\left\{ -\tilde{y}_{1}+\tilde{S}\left( -1\right) ^{n}n%
\left[ \left( \tilde{y}_{1}\right) ^{2}-3\tilde{y}_{2}\right] ^{1/2}\right\}
~,~~~n=1,2~.  \label{EvEq2}
\end{equation}%
Our next step is to then assume again that the two coefficients $\tilde{y}%
_{1}\equiv y_{1}\left( \ell +1\right) $ and $\tilde{y}_{2}\equiv y_{2}\left(
\ell +1\right) $ evolve according to the \textit{solvable} \textit{%
discrete-time} system (\ref{2y12tilde}). By proceeding in close analogy with
the previous treatment---i. e., by replacing in the right-hand sides of (\ref%
{EvEq2}) the variables $\tilde{y}_{1}$ and $\tilde{y}_{2}$ via the evolution
equations (\ref{2y12tilde}) and then in the right-hand sides of the
resulting equations $y_{1}$ and $y_{2}$ via (\ref{y123x12})---we thereby
obtain the following

\textbf{Proposition 2.3}. The solution of the initial-values problem for the
following system of \textit{discrete-time} evolution equations
\begin{subequations}
\label{2EvEq3}
\begin{equation}
\tilde{x}_{n}=\left( 1/2\right) \left\{ -\alpha \left[ -\left(
2x_{1}+x_{2}\right) \right] ^{k+1}+\left( -1\right) ^{n}S\Delta \right\}
~,~~~n=1,2~,  \label{2EvEq3a}
\end{equation}%
\begin{eqnarray}
\Delta &=&\left\{ \alpha ^{2}\left[ -\left( 2x_{1}+x_{2}\right) \right]
^{2\left( k+1\right) }-3\beta ^{2}\left( x_{1}\right) ^{2}x_{2}\left[
-\left( 2x_{1}+x_{2}\right) \right] ^{q}\right.  \notag \\
&&\left. -3\gamma \left[ -\left( 2x_{1}+x_{2}\right) \right] ^{r}\right\}
^{1/2}~,
\end{eqnarray}%
is provided by the $2$ zeros $x_{1}\equiv x_{1}\left( \ell \right) $ and $%
x_{2}\equiv x_{2}\left( \ell \right) $ of the polynomial (\ref{2Pol3})---i.
e., by the formulas (\ref{2Solx12y12})---with the coefficients $y_{1}\equiv
y_{1}\left( \ell \right) $ and $y_{2}\equiv y_{2}\left( \ell \right) $ given
by the formulas (\ref{2y12}) where of course now (see (\ref{y123x12})) $%
y_{1}\left( 0\right) =-2x_{1}\left( 0\right) -x_{2}\left( 0\right) $ and $%
y_{2}\left( 0\right) =x_{1}\left( 0\right) \left[ x_{1}\left( 0\right)
+2x_{2}\left( 0\right) \right] $.

Note that this system features the same kind of $2$-fold ambiguity as
discussed in the previous \textbf{Section 1} (see after eq. (\ref{1xndotk}%
)). \textbf{$\blacksquare $}

However, a "defect" of this \textit{solvable} system is the appearance in
the right-hand side of its \textit{discrete-time} equations of motion (\ref%
{2EvEq3}) of a square root; but this "defect" can now be eliminated by
restricting the parameters $q$ and $r$ to satisfy the condition (\ref{2qrk}%
)---just the same condition that allows to replace the solution (\ref{2y12})
with the \textit{more explicit} solution (\ref{2y12Spec})---and by moreover
replacing the assignments (\ref{gammaCase1}) with the following assignments
\end{subequations}
\begin{equation}
\gamma =3\left( a^{2}-b^{2}\right) ~,~~~\alpha =3a~,~~~\beta =3b~.
\label{gammaCase2}
\end{equation}%
It is indeed easily seen that there thereby holds the following

\textbf{Proposition 2.4}. The solution of the initial-values problem for the
system of \textit{discrete-time} evolution equations (\ref{1xndotk}) is
provided by the $2$ distinct \textit{zeros} $x_{1}\equiv x_{1}\left( \ell
\right) $ and $x_{2}\equiv x_{2}\left( \ell \right) $ of the polynomial (\ref%
{2Pol3})---i. e., by the formulas (\ref{2Solx12y12})---with the \textit{%
coefficients} $y_{1}\equiv y_{1}\left( \ell \right) $ and $y_{2}\equiv
y_{2}\left( \ell \right) $ given by the formulas (\ref{2y12Spec}) where of
course now (see (\ref{y123x12})) $y_{1}\left( 0\right) =-2x_{1}\left(
0\right) -x_{2}\left( 0\right) $ and $y_{2}\left( 0\right) =x_{1}\left(
0\right) \left[ x_{1}\left( 0\right) +2x_{2}\left( 0\right) \right] $.

Let us again emphasize that, for each value of the \textit{discrete-time} $%
\ell $, this prescription yields $2$ different solutions, say the $2$
different pairs $x_{1}^{\left( +\right) }\left( \ell \right) ,$ $%
x_{2}^{\left( +\right) }\left( \ell \right) $ and $x_{1}^{\left( -\right)
}\left( \ell \right) ,$ $x_{2}^{\left( -\right) }\left( \ell \right) $.
\textbf{$\blacksquare $}

This \textbf{Proposition 2.4 }corresponds to the second \textit{main} result
reported in \textbf{Section 1}.

\bigskip

\section{Additional developments}

An important issue is the possibility to generalize the \textit{%
algebraically solvable} systems treated in the previous \textbf{Section 2}%
---which feature the $2$ \textit{arbitrary} (possibly \textit{complex) }%
parameters $a$ and $b$---to more general analogous models involving more
free parameters. Following the treatment given in \cite{CP2019a}, let us
outline how this can be done for the system (\ref{1xndotk}).

The procedure is to introduce the simple invertible change of dependent
variables
\begin{subequations}
\label{zx}
\begin{equation}
z_{1}=A_{11}x_{1}+A_{12}x_{2}~,~~~z_{2}=A_{21}x_{1}+A_{22}x_{2}~,
\end{equation}%
\begin{equation}
x_{1}=\left( A_{22}z_{1}-A_{12}z_{2}\right) /D~,~~~x_{2}=\left(
-A_{21}z_{1}+A_{11}z_{2}\right) /D~,
\end{equation}%
\begin{equation}
D=A_{11}A_{22}-A_{12}A_{21}~,
\end{equation}%
where of course the $4$ parameters $A_{nm}$ ($n=1,2;~m=1,2$) are $4$,
\textit{a priori arbitrary}, time-independent constants (the restriction to
time-independent constants is because we prefer in this paper to focus on
\textit{autonomous} systems of \textit{discrete-time} evolutions).

It is then a matter of simple algebra to obtain the---of course \textit{%
algebraically solvable}---evolution equations satisfied by the dependent
variables $z_{1}\equiv z_{1}\left( \ell \right) $ and $z_{2}\equiv
z_{2}\left( \ell \right) $:
\end{subequations}
\begin{subequations}
\begin{eqnarray}
\tilde{z}_{n} &=&D^{-\left( k+1\right) }\left[ \left( 2A_{22}-A_{21}\right)
z_{1}+\left( A_{11}-2A_{12}\right) z_{2}\right] ^{k}\cdot  \notag \\
&&\cdot \left[ A_{n1}f_{1}\left( z_{1},z_{2};k\right) +A_{n2}f_{2}\left(
z_{1},z_{2};k\right) \right] ~,~~~n=1,2~,
\end{eqnarray}%
\begin{equation}
f_{n}\left( z_{1},z_{2};k\right) =\left( \theta _{k;2,n;1}A_{22}-\theta
_{k;1,n;1}A_{21}\right) z_{1}+\left( \theta _{k;1,n;0}A_{11}+\theta
_{k;2,n;0}A_{12}\right) z_{2}~,  \label{f12}
\end{equation}%
~%
\begin{equation}
\theta _{k;n_{1},n_{2};n}=\left( -1\right) ^{k}n_{1}a+\left( -1\right)
^{n}n_{2}Sb~.  \label{theta}
\end{equation}

Let us also display these equations in the---possibly more relevant to
applicative contexts---special cases with $k=\pm 1$.

For $k=1$:
\end{subequations}
\begin{subequations}
\begin{equation}
\tilde{z}_{n}=a_{n1}\left( z_{1}\right) ^{2}+a_{n2}\left( z_{2}\right)
^{2}+a_{n3}z_{1}z_{2}~,~~~n=1,2~,
\end{equation}%
\begin{equation}
a_{n1}=D^{-2}\lambda _{2}\left( \eta _{1}A_{n1}+\eta _{2}A_{n2}\right)
~,~~~n=1,2~,
\end{equation}%
\begin{equation}
a_{n2}=D^{-2}\lambda _{1}\left( \eta _{3}A_{n1}+\eta _{4}A_{n2}\right)
~,~~~n=1,2~,
\end{equation}%
\begin{equation}
a_{n3}=D^{-2}\left[ \lambda _{1}\left( A_{n1}\eta _{11}+A_{n2}\eta
_{21}\right) +\lambda _{2}\left( A_{n1}\eta _{12}+A_{n2}\eta _{22}\right) %
\right] ~,~~~n=1,2~,
\end{equation}%
\begin{equation}
\lambda _{n}=\left( -1\right) ^{n}\left( 2A_{n2}-A_{n1}\right) ~,~~~n=1,2~,
\label{landa12}
\end{equation}%
\begin{equation}
\eta _{n1}=-\theta _{1;1,n;n}A_{21}-\theta _{0;2,n;n+1}A_{22}~,~~~n=1,2~,
\end{equation}%
\begin{equation}
\eta _{n2}=-\theta _{0;1,n;n}A_{11}+\theta _{0;2,n;n+1}A_{12}~,~~~n=1,2~.
\end{equation}

For $k=-1$:
\end{subequations}
\begin{subequations}
\begin{equation}
\tilde{z}_{n}=\frac{A_{n1}f_{1}\left( z_{1},z_{2};-1\right)
+A_{n2}f_{2}\left( z_{1},z_{2};-1\right) }{\lambda _{1}z_{1}+\lambda
_{2}z_{2}}~,~~~n=1,2~,
\end{equation}%
with $f_{n}\left( z_{1},z_{2};-1\right) $ and $\lambda _{n}$ ($n=1,2$)
defined as above, see (\ref{f12}), (\ref{theta}) and (\ref{landa12}).

An alternative generalization of this approach is based on the replacement
of the relations (\ref{y12x12}) and (\ref{y123x12}) and their generalization
via (\ref{zx}) with the following more general relations:
\end{subequations}
\begin{equation}
y_{1}=B_{1}z_{1}+B_{2}z_{2}~,~~~y_{2}=C_{1}\left( z_{1}\right)
^{2}+C_{2}\left( z_{2}\right) ^{2}+C_{3}z_{1}z_{2}~.  \label{yz}
\end{equation}%
Note that these relations involve the $5$ \textit{a priori arbitrary}
parameters $B_{n}$ and $C_{j}$ ($n=1,2$; $j=1,2,3$), and that they are
easily inverted:
\begin{subequations}
\begin{equation}
z_{1}=\frac{-g\pm \Gamma }{2f}~,~~~z_{2}=\frac{y_{1}-B_{1}z_{1}}{B_{2}}~,
\end{equation}%
\begin{equation}
f=\frac{\left( B_{2}\right) ^{2}C_{1}+\left( B_{1}\right)
^{2}C_{2}-C_{3}B_{1}B_{2}}{\left( B_{2}\right) ^{2}}~,~~~g=\frac{\left(
B_{2}C_{3}-2B_{1}C_{2}\right) y_{1}}{\left( B_{2}\right) ^{2}}~,
\end{equation}

\begin{equation}
\Gamma ^{2}=g^{2}-4fh~,~~~h=\frac{C_{2}\left( y_{1}\right) ^{2}-\left(
B_{2}\right) ^{2}y_{2}}{\left( B_{2}\right) ^{2}}~.
\end{equation}

Starting from these formulas, and proceeding in close analogy with the
treatment provided above---which involve of course the assumption that the
quantities $z_{n}$ and $y_{n}$ ($n=1,2$) are $\ell $-dependent while the
parameters $B_{n}$ and $C_{j}$ ($n=1,2$; $j=1,2,3$) are $\ell $-independent,
and moreover that the quantities $y_{n}$ evolve according to the \textit{%
solvable discrete-time} evolution equations (\ref{2y12tilde})---one arrives
at the following---of course, also \textit{solvable---}evolution equations
for the quantities $z_{n}\equiv z_{n}\left( \ell \right) $:
\end{subequations}
\begin{subequations}
\begin{equation}
\tilde{z}_{1}=d\left( B_{1}z_{1}+B_{2}z_{2}\right) ^{k}\left[ \alpha
g_{1}\left( B_{1}z_{1}+B_{2}z_{2}\right) +S\beta B_{2}\left(
g_{2}z_{1}+g_{3}z_{2}\right) \right] ~,
\end{equation}%
\begin{eqnarray}
&&\tilde{z}_{2}=\left( B_{2}\right) ^{-1}\left( B_{1}z_{1}+B_{2}z_{2}\right)
^{k}\cdot  \notag \\
&&\cdot \left[ \alpha \left( 1-B_{1}dg_{1}\right) \left(
B_{1}z_{1}+B_{2}z_{2}\right) -B_{1}dS\beta B_{2}\left(
g_{2}z_{1}+g_{3}z_{2}\right) \right] ~;
\end{eqnarray}%
\begin{equation}
d=\left( 1/2\right) \left[ \left( B_{1}\right) ^{2}C_{2}+\left( B_{2}\right)
^{2}C_{1}-B_{1}B_{2}C_{3}\right] ^{-1}~,  \label{d}
\end{equation}%
\begin{equation}
g_{1}=2B_{1}C_{2}-B_{2}C_{3}~,~~~g_{2}=2B_{2}C_{1}-B_{1}C_{3}~,~~~g_{3}=B_{2}C_{3}-2B_{1}C_{2}~.
\label{g123}
\end{equation}%
Note again the presence---in the right-hand side of the first of these
formulas---of the sign $S\equiv S\left( \ell \right) ,$ and that these
equations---involving no square roots in their right-hand sides---have been
obtained thanks to the assignments (\ref{2qrk}) and by setting (in place of (%
\ref{gammaCase1}) respectively (\ref{gammaCase2}))%
\begin{equation}
\gamma =\frac{\left[ \left( C_{3}\right) ^{2}-4C_{1}C_{2}\right] \left(
\beta ^{2}-\alpha ^{2}\right) }{4\left[ \left( B_{1}\right) ^{2}C_{2}+\left(
B_{2}\right) ^{2}C_{1}-B_{1}B_{2}C_{3}\right] }~.
\end{equation}

The special cases of these equations corresponding to the assignments $k=1$
and $k=-1$ are also worth \textit{explicit} display:
\end{subequations}
\begin{subequations}
\begin{equation}
\tilde{z}_{1}=d\left( B_{1}z_{1}+B_{2}z_{2}\right) ^{k}\left[ \alpha
g_{1}\left( B_{1}z_{1}+B_{2}z_{2}\right) +S\beta B_{2}\left(
g_{2}z_{1}+g_{3}z_{2}\right) \right] ~,
\end{equation}%
\begin{eqnarray}
&&\tilde{z}_{2}=\left( B_{2}\right) ^{-1}\left( B_{1}z_{1}+B_{2}z_{2}\right)
^{k}\cdot  \notag \\
&&\cdot \left[ \alpha \left( 1-B_{1}dg_{1}\right) \left(
B_{1}z_{1}+B_{2}z_{2}\right) -B_{1}dS\beta B_{2}\left(
g_{2}z_{1}+g_{3}z_{2}\right) \right] ~;
\end{eqnarray}%
here of course $d$ respectively $g_{j}$ ($j=1,2,3$) are defined in terms of
the $5$ parameters $B_{n}$ and $C_{j}$ ($n=1,2$; $j=1,2,3$) by (\ref{d})
respectively (\ref{g123}).

Assigning \textit{solvable} evolutions to $y_{1}$ and $y_{3}$ or $y_{2}$ and
$y_{3}$ (rather than to $y_{1}$ and $y_{2}$; in the case of the \textit{%
third-degree} polynomial (\ref{Pol3})) are possible further developments,
but we postpone the relevant treatments to future papers.

\bigskip

\section{Acknowledgements}

Both authors like to thank Piotr Grinevich, Paolo Santini and Nadezda
Zolnikova for very useful suggestions. FP likes to thank the Physics
Department of the University of Rome\ "La Sapienza" for the hospitality from
February 2018 to April 2019 (during her sabbatical), when the results
reported in this paper were obtained.

\end{subequations}

\end{document}